\documentstyle[aps,preprint,epsfig]{revtex}
\newcommand{\beq}{\begin{equation}}
\newcommand{\eeq}{\end{equation}}
\newcommand{\bdis}{\begin{displaymath}}
\newcommand{\edis}{\end{displaymath}}
\newcommand{\bea}{\begin{eqnarray}}
\newcommand{\eea}{\end{eqnarray}}
\newcommand{\barr}{\begin{array}}
\newcommand{\earr}{\end{array}}

\newcommand{\equ}[1]{(\protect\ref{#1})}
\newcommand{\p}{\psi}
\newcommand{\pb}{\bar{\psi}}
\newcommand{\f}{\phi}
\newcommand{\fb}{\bar{\phi}}
\newcommand{\la}{\lambda}

\begin{document}
\draft
\title{Field theory of absorbing phase transitions 
with a non-diffusive conserved field}
 
\author{Romualdo Pastor-Satorras$^{(1,2)}$  
and Alessandro Vespignani$^{(2)}$}

\address{1) Dept. de F{\'\i}sica Fonamental, Facultat de
  F{\'\i}sica, Universitat de Barcelona, Av. Diagonal 647,
  08028 Barcelona, Spain \\
  2) The Abdus Salam International Centre for Theoretical Physics
  (ICTP),
  P.O. Box 586, 34100 Trieste, Italy\\}
   \date{\today} 

\maketitle

\begin{abstract}
  We investigate the critical behavior of a reaction-diffusion system
  exhibiting a continuous absorbing-state phase transition. The
  reaction-diffusion system strictly conserves the total density of
  particles, represented as a non-diffusive 
  conserved field, and allows an
  infinite number of absorbing configurations.  Numerical results show
  that it belongs to a wide universality class that also includes
  stochastic sandpile models. We derive microscopically the field
  theory representing this universality class.
\end{abstract}

\pacs{PACS numbers: 64.60.Ht, 05.70.Ln, 05.65.+b, 05.50.+q}

The directed percolation (DP)\cite{reviews} universality class is
recognized as the canonical example of the critical behavior in the
transition from an active to a single absorbing state.  This
universality class appears to be very robust with respect to
microscopic modifications, and non-DP behavior emerges only in the
presence of additional symmetries, such as in the case of symmetric
absorbing states\cite{cardytauber96}, long-range
interactions\cite{lr99}, or infinitely many absorbing
states\cite{many}.

Recently a new universality class of absorbing-state phase transitions
(APT)\cite{reviews} coupled to a non-diffusive conserved field has
been identified\cite{rpv00}.  This class characterizes the critical
behavior of several models showing absorbing transitions with a
dynamics that strictly conserves the density of particles, that is
represented by a conserved static (non-diffusive) field. The models
are tuned to criticality by varying the particle density in the
initial state, and exhibit an infinite number of absorbing states.
This universality class is particularly interesting because it
embraces also the large group of stochastic sandpile
models\cite{jenssen98} (and in particular the Manna model
\cite{mannamodel}) which are the prototypical examples that illustrate
the ideas of self-organized criticality (SOC) \cite{btw}. These are
driven dissipative models in which sand (or energy) is injected into
the system and dissipated through the boundaries, leading eventually
to a stationary state. In the limit of infinitesimally slow external
driving, the systems approach a critical state characterized by an
avalanche-like response.  Recently it has been pointed out that this
critical state is equivalent to the absorbing phase transition present
in the {\em fixed energy} case, that is, in automata that consider the
same microscopic rules defining the sandpile dynamics, but without
driving nor dissipation\cite{fes,bigfes,pacz}.

The numerical evidence for the existence of such a general
universality class\cite{rpv00} is corroborated by the observation
that all the models analyzed share the same structure and basic
symmetries; namely, a conserved and static non-critical field
dynamically coupled to a non-conserved order parameter field,
identified as the density of the active particles.  These observations
have led to the conjecture that, in absence of additional symmetries,
{\em all stochastic models with an infinite number of absorbing states
  in which the order parameter evolution is coupled to a non-diffusive
  conserved field define a unique universality class} \cite{rpv00}.

In this Letter, we study the non-diffusive field limit for the two species
reaction-diffusion model introduced in Ref.~\cite{wij98} (see also
Ref.~\cite{kree}). In this limit the model has a phase transition with
infinitely many absorbing states and it conserves the total number of
particles, that is associated to a non-diffusive conserved field. We present
extensive numerical simulations of the model in two and three
dimensions, and determine the full set of critical exponents. The
obtained values are compatible with the new universality class
conjectured in Ref.\cite{rpv00}.  This definitely shows the existence
of a very broad universality class that includes reaction-diffusion
processes, stochastic sandpile models, and lattice gases with the same
symmetry properties. For the present reaction-diffusion model, it is
possible to derive microscopically a field theory (FT) description.
The resulting action and Langevin equations exhibit the basic
symmetries that characterize this universality class, and represent
the first microscopic derivation of a FT for sandpile models.
Notably, the resulting FT description recovers a phenomenological
Langevin approach proposed for stochastic sandpiles\cite{fes,bigfes}.
The analysis provided here is a very promising path for a coherent
description of several nonequilibrium critical phenomena now
rationalized in a single universality class.

We consider the two-component reaction-diffusion process identified by
the following set of reaction equations:
\begin{eqnarray}
  B &\to& A \quad \mbox{\rm with rate $k_1$}. \label{reaction1}\\
  B + A &\to& 2 B  \quad \mbox{\rm with rate $k_2$},\label{reaction2} 
\end{eqnarray}
In this system, $B$ particles diffuse with diffusion rate $D_B\equiv
D$, and A particles {\em do not diffuse}, that is, $D_A=0$.  This
corresponds to the limit $D_A\to 0$ of the model introduced in
Ref.~\cite{wij98}.  From the rate Eqs.~\equ{reaction1}
and~\equ{reaction2}, it is clear that the dynamics conserves the total
density of particles $\rho=\rho_A+\rho_B$, where $\rho_i$ is the
density of component $i=A, B$. In this conserved reaction-diffusion
model, the only dynamics is due to $B$ particles, that we can identify
as {\em active} particles. $A$ particles do not diffuse and cannot
generate spontaneously $B$ particles.  More specifically, $A$
particles can only move via the motion of $B$ particles that later on
transform into $A$ because of Eq.~(\ref{reaction2}). In absence of $B$
particles, $\rho_A$ can be considered a static field.  This implies
that any configuration devoid of $B$ particles is an absorbing state
in which the system is trapped forever.

It is easy to see \cite{wij98} that the reaction-diffusion process
defined by Eqs.~(\ref{reaction1}) and (\ref{reaction2}) exhibits a
phase transition from an active to an absorbing phase for a
non-trivial value of the total particle density $\rho=\rho_c$, which
acts as the control parameter.  The critical value $\rho_c$ depends
upon the reaction rates $k_1, k_2$.  The nature of this phase
transition (whether it is first or second order) for $D_A\neq0$
appears to be determined by the ratio between $D_B$ and
$D_A$\cite{wij98}; the static field case ($D_A=0$), on the other hand,
has never been explored.  It is clear that the static field conserved
reaction-diffusion (SFCRD) model allows, for any density $\rho$, an
infinite number (in the thermodynamic limit) of absorbing
configurations, in which there are no $B$ particles.  This is the key
difference with respect to the case in which $D_A\neq 0$, as in
Ref.\cite{wij98}. In the latter case a configuration devoid of $B$
particles consists of many diffusing $A$ particles, jumping from site
to site. In the long run, all particles can visit all sites, and
therefore, in a statistical sense, all configuration with a fixed
number of $A$'s are equivalent and the absorbing state can be
considered statistically unique \cite{mendes94}.

The present SFCRD model seems to possess all the required symmetry
(stochastic dynamics, many absorbing states, static conserved field)
for being part of the universality class conjectured in
Ref.\cite{rpv00}.  In order to test this possibility, we have
performed numerical simulations of the model in a $d$-dimensional
hypercubic lattice with $N=L^d$ sites. Each site can store any number
of $A$ and $B$ particles; that is, our model can be represented by
bosonic variables. Initial conditions are generated by randomly
placing $N \rho_A^{(0)}$ particles $A$ and $N \rho_B^{(0)}$ particles
$B$, corresponding to a particle density
$\rho=\rho_A^{(0)}+\rho_B^{(0)}$.  The results are independent of the
particular initial ratio $\rho_A^{(0)}/\rho_B^{(0)}$, apart from very
early time transients.  The dynamics proceeds in parallel. Each time
step, we update the lattice according to the following rules. a)
Diffusion: on each lattice site, each $B$ particles moves into a
randomly chosen nearest neighbor site.  b) After all sites have been
updated for diffusion, we perform the reactions: i) On each lattice
site, each $B$ particle is turned into an $A$ particle with
probability $r_1$. ii) At the same time, each $A$ particle becomes a
$B$ particle with probability $1-(1-r_2)^{n_B}$, where ${n_B}$ is the
total number of $B$ particles in that site. This corresponds to the
average probability for an $A$ particle of being involved in the
reaction~(\ref{reaction2}) with any of the $B$ particles present on
the same site.  The probabilities $r_1$ and $r_2$ are related to the
reaction rates $k_1$ and $k_2$ defined in Eqs.~\equ{reaction1} and
\equ{reaction2}.  The order parameter of the system is $\rho_B$,
measuring the density of dynamical entities.  For small initial
densities $\rho$, the system will very likely fall into an absorbing
configurations with only frozen $A$ particles. For large densities,
the system reaches a stationary active state with $\rho_B\neq 0$.

As we vary $\rho$, the system exhibits a continuous transition separating
an absorbing phase from an active phase at a critical point $\rho_c$.
The order parameter is null for $\rho<\rho_c$, and follows a power law
$\rho_B\sim (\rho-\rho_c)^\beta$, for $\rho\geq\rho_c$.  The system
correlation length $\xi$ and time $\tau$, that define 
the exponential relaxation of space and time correlation functions, both 
diverge as $\rho\to\rho_c$\cite{reviews}. 
In the critical region the system is characterized by a
power law behavior, namely $\xi\sim |\rho-\rho_c|^{-\nu_\perp}$ and
$\tau\sim |\rho-\rho_c|^{-\nu_\parallel}$. The dynamical critical
exponent is defined as $\tau\sim\xi^z$, with
$z=\nu_\parallel/\nu_\perp$.  These exponents fully determine the
critical behavior of the stationary state of the model (see 
Ref.~\cite{reviews}).

\begin{figure}[t]
    \centerline{\epsfig{file=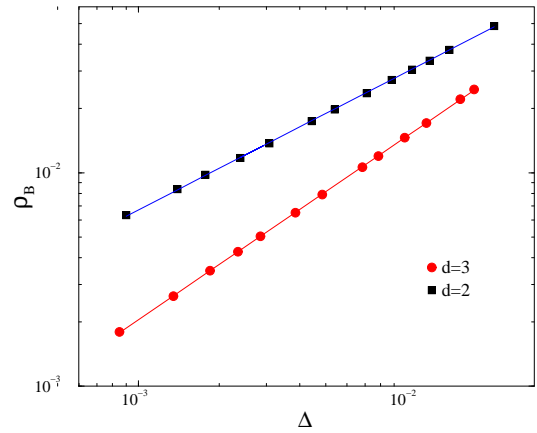, width=7cm}}
  \caption{Order parameter behavior (stationary density of $B$ particles)
as a function of $\Delta=\rho-\rho_c$ for the reaction diffusion 
model in $d=2$ and $3$. The slope of the straight lines is $\beta=0.65$ 
in $d=2$ and $\beta=0.86$ in $d=3$.}
  \label{op}
\end{figure}

We have studied the steady-state properties of the model in $d=2$ and
$3$, by performing numerical simulations for systems with size ranging
up to $L=512$ and $L=125$, respectively. Averages were performed over
$10^4-10^5$ independent initial configurations. The values considered
for the rates $r_i$ are $r_1=0.1$ and $r_2=0.5$ in $d=2$, and
$r_1=0.4$ and $r_2=0.5$ in $d=3$.  From the finite-size scaling
analysis for absorbing phase transitions\cite{reviews} we obtain the
critical point ($\rho_c=0.3226(1)$ in $d=2$ and $\rho_c=0.95215(15)$ 
in $d=3$) and the complete set of critical exponents. 
A detailed presentation of
these results will be reported elsewhere.  In
Fig.~\ref{op} we show as an example the order parameter behavior with
respect to the control parameter $\Delta=\rho-\rho_c$, from
which it is possible to calculate directly the $\beta$ exponent. The
results obtained in $d=2$ and $3$ are reported in Tables I and II and
compared with the Manna sandpile model in the respective dimension.

In APT it is possible to obtain more information on the critical state
by studying the evolution (spread) of activity in systems which start
close to an absorbing configuration \cite{grassberger79}. In each {\em
  spreading} simulation, a small perturbation is added to an absorbing
configuration. It is then possible to measure the spatially integrated
activity $N(t)$, averaged over all runs, and the survival probability
$P(t)$ of the activity after $t$ time steps.  Only at the critical
point we have power law behavior for these magnitudes.  In the case of
many absorbing states, the choice of the initial absorbing state is
not unique\cite{jensen93b}. There are several methods to perform
spreading exponents in this case, and we have followed the technique
outlined in Ref.\cite{rpv00}. This procedure amounts to the study of
critical spreading with the so-called ``natural initial conditions''
at $\rho=\rho_c$ \cite{jensen93b}.  The probability distribution
$P_s(s)$ of having a spreading event involving $s$ sites, as well as
the the quantities $N(t)$ and $P(t)$ can thus be measured.  At
criticality, the only characteristic length is the system size $L$,
and we can write the scaling forms $P_s(s)=s^{-\tau_s}h(s/L^D)$,
$N(t)=t^\eta f(t/L^z)$, and $P(t)=t^{-\delta}g(t/L^z)$
\cite{grassberger79}. The scaling functions $f(x)$, $g(x)$ and $h(x)$
are decreasing exponentially for $x\gg1$, and we have considered that
the spreading characteristic time and size are scaling as $L^z$ and
$L^D$, respectively.  In this case simulations were performed for
systems of size up to $L=1024$ in $d=2$ and $L=200$ in $d=3$,
averaging over at least $5\times10^6$ spreading experiments.  The new
scaling exponents $\tau_s, D, \delta$, and $\eta$ are measured using
the now standard moment analysis technique \cite{moments,mannasim},
and a consistency check is executed by performing a data collapse
analysis.  The resulting exponents are summarized in Tables I and II.
As a further consistency check of our results, we have checked that
our exponents fulfill all scaling and hyper-scaling relations in
standard APT.  Despite the apparent diversity in the dynamical rules,
we can safely include the SFCRD model and the Manna model in the same
universality class.  The reported numerical values provide striking
evidence for a single universality class, and a further check of the
conjecture in Ref.\cite{rpv00}.

From a theoretical point of view, the SFCRD allows the microscopic
construction of a field theory description that will represent also
the critical behavior of all models belonging to the same universality
class. The construction of the FT follows the standard steps outlined
in the work of Doi, Peliti, and also in Lee and Cardy\cite{ft}.  This
technique consists in recasting the master equation implicit in
Eqs.~\equ{reaction1} and \equ{reaction2} into a ``second quantized
form'' via a set of creation and annihilation bosonic operators for
particles $A$ and $B$ on each site.  At this point it is possible to
map the solution of the master equation in a path integral on the
density fields weighted by the exponential of a functional action
$S$\cite{ft}.  In our case, we can quote the elegant results of
Ref.\cite{wij98}, just considering that we have $D_A=0$. The action of
the FT is thus
\begin{eqnarray}
  S&=&\int dx dt \left\{ \pb [\partial_t + 
        (r-D \nabla^2)]\p + \fb [\partial_t \f  -  \la  \nabla^2\p]
        \right. \nonumber \\  
        &+&   u_1 \pb \p (\p - \pb) +  u_2 \pb \p (\f+\fb) 
        \nonumber  \\   
        &+&  \left.  v_1  \pb^2 \p^2  + v_2 \pb \p (\p \fb-\pb \f) 
    + v_3 \pb \p \f \fb\right\},
\label{eq:actionreal}
\end{eqnarray}
where $\p$ and $\f$ are auxiliary fields, defined such that their
average values coincide with the average density of $B$ particles and
the total density of particles, respectively, $\pb$ and $\fb$ are
response fields, and the coupling
constants are related to the reaction rates $k_i$.  Namely, $D$
represents the diffusion coefficient of $B$ particles, $\lambda$ is
initially also proportional to $D$, and $r$ is the critical parameter
that plays the role of the mass in the FT and is related to the
difference of the total density with respect to the critical density
$\rho_c$.  The critical point in the mean-field theory is of course at
$r_c=0$.  By standard power-counting analysis, one realizes that the
reduced couplings $u_i/D$ have critical dimension $d_c^{(1)}=4$, while
the couplings $v_i/D$ have on their part $d_c^{(2)}=2$. This means
that when applying the renormalization group (RG) and performing a
perturbative expansion around the critical dimension $4$, one could in
principle drop all the couplings $v_i$\cite{notev}. The critical
parameter of this theory is the density of active sites $\p$, while
$\f$ serves just to propagate interactions. We can exploit some
symmetry considerations of the FT to relate the physics of the system
to the corresponding analytical description. In fact, by neglecting
irrelevant terms in the power counting analysis, the action
\equ{eq:actionreal} is invariant under the shift transformation
\begin{equation}
  \f\to \f'=\f + \delta, \qquad r \to r'=r-u_2 \delta,
\label{eq:shift}
\end{equation}
where $\delta$ is any constant.  This symmetry has a very intuitive
meaning: If we increase everywhere the density of the system by an
amount $\delta$, we must be closer to the critical point by an amount
proportional to $\delta$.  In other words, this symmetry represents
the conserved nature of the system.  It is also interesting to write
the set of corresponding Langevin equations (up to irrelevant terms) 
by integrating out the response fields $\pb, \fb$ in the 
action $S$:
\begin{eqnarray}
  \partial_t \p &=& D \nabla^2 \p - r \p - 
  u_1 \p^2 -  u_2 \p \f
  + \eta_\p, \\
  \partial_t \f &=& \la  \nabla^2 \p + \eta_\f.
\end{eqnarray}
Here, $\eta_\p$ and $\eta_\f$ are noise term with zero mean and
correlations $ \left< \eta_\p(x,t) \eta_\p(x',t') \right> = 2 u_1
\p(x,t)\delta(x-x') \delta(t-t')$, $\left< \eta_\p(x,t) \eta_\f(x',t')
\right>= - u_2 \p(x,t)\delta(x-x')\delta(t-t')$ and $ \left<
  \eta_\f(x,t) \eta_\f(x',t') \right>=0$.  The noise terms have a
multiplicative nature \cite{rft}, that is the standard form in APT.
Note that the $v_i$ couplings contribute to the noises correlations
with higher order terms.  These equations have a very clear physical
interpretation. The field $\f$ is conserved and static, i.e., it only
diffuses via the activity of $B$ particles, represented by the field
$\p$\cite{notecon}.  On its turn, the field $\p$ is locally coupled to
the field $\f$, but is non-conserved.  Noticeably, this set of
equations recovers the Langevin description that has been proposed on
a phenomenological level for stochastic sandpiles\cite{fes,bigfes},
with the extra infromation of the cross-correlation term $\left<
  \eta_\p \eta_\f \right>$ .  Indeed, the sandpile model has the same
basic symmetries of the present reaction-diffusion model, once the
local density field $\rho$ is replaced by the local sand-grain
(energy) density and the order parameter is identified with the
density of toppling sites field\cite{fes,bigfes}.  It is then natural
to expect that the very same basic structure is reflected in a unique
theoretical description.  This observation substantiates the existence
of a common universality class that embraces stochastic sandpiles,
conserved lattice gases and reaction diffusion systems with many
absorbing states.

The complete RG analysis of the field theory would allow to extract
estimates for the critical exponents to compare with simulations in
$d=2$ and $3$. Unfortunately, several severe technical problems are
encountered in this case. In general, as pointed out in Ref.\cite{wij98}, 
the couplings $v_i$
become relevant and should be taken into account in the RG analysis.
The importance of the couplings $v_i$ can be argued by the change of
the energy shift symmetry form, Eq.~\equ{eq:shift}, in the case of the full
action, Eq.~(\ref{eq:actionreal}).  Second, and more important, is the
presence of the singular bare propagator for the field $\f$, that
cannot be regularized by adding a mass term $m^2 \f \fb$, since it
will obviously break the symmetry \equ{eq:shift}. This singular
propagator gives rise to divergences in the RG perturbative
expansions, and the results of Ref.\cite{wij98} cannot be extended
``{\em tout-court}'' to the limit $D_A\to 0$. In particular, some
Feynman diagrams in the $\epsilon$-expansion presented in Refs.
\cite{wij98,kree} are proportional to $1/D_A$. Hence, the limit
$D_A\to0$ in the theory with $D_A\neq0$ is non-analytic; any
infinitesimal amount of diffusion in the energy field renormalizes to
a finite value, and definitely changes the universality class of the
model.  Work is in progress to provide a suitable regularization that
will allow an $\epsilon$-expansion calculation of the FT critical
exponents.

This work has been supported by the European Network under Contract
No.~ERBFMRXCT980183. We thank D. Dhar, R. Dickman, P. Grassberger,
H. J. Hilhorst, 
M.A. Mu{\~n}oz, F.~van~Wijland and S. Zapperi, for helpful comments 
and discussions.

\newpage
\begin{table}[t]
\begin{tabular}{lccccc}
  &  \multicolumn{5}{c}{Steady state exponents} \\
  \cline{2-6}
  & $\beta$ & $\beta/\nu_\perp$ & $\nu_\perp$ & $z$ & $\nu_\parallel$ \\
  \hline
  SFCRD &   $0.65(1)$ & $0.78(2)$ & $0.83(3)$ &$ 1.55(5)$& $1.29(8)$\\
  Manna & $0.64(1)$ & $0.78(2)$ & $0.82(3)$ &$1.57(4)$&$1.29(8)$ \\ 
    \hline \hline
  &  \multicolumn{5}{c}{Spreading exponents} \\
  \cline{2-6}
  & $\tau_s$  & $D$ & $z$ & $\eta$  &  $\delta$ \\
  \hline
  SFCRD   & $1.27(1)$ & $2.75(1)$ & $1.54(2)$ & $0.29(2)$ &
  $0.50(2)$ \\
    Manna & $1.28(1)$ & $2.76(1)$ & $1.55(1)$ & $0.30(3)$
  & $0.48(2)$ 
 \end{tabular}
\caption{Critical exponents for spre\-ading and ste\-ady state
  experiments in $d=2$. Figures in parenthesis indicate the statistical 
  uncertainty in the last digit. Manna exponents from
  Refs.~\protect\cite{rpv00,bigfes,mannasim,mfdp1,mfdp2}. }
\label{tableI}
\end{table}

\begin{table}[t]
\begin{tabular}{lccccc}
  &  \multicolumn{5}{c}{Steady state exponents} \\
  \cline{2-6}
  & $\beta$ & $\beta/\nu_\perp$ & $\nu_\perp$ &$z$ & $\nu_\parallel$ \\
  \hline
  SFCRD &   $0.86(2)$ & $1.39(4)$ & $0.62(3)$ &$ 1.80(5)$& $1.12(8)$\\
  Manna & $0.84(2)$ & $1.40(2)$ & $0.60(3)$ &$1.80(5)$& $1.08(8)$\\ 
    \hline \hline
  &  \multicolumn{5}{c}{Spreading exponents} \\
  \cline{2-6}
  & $\tau_s$  & $D$ & $z$ & $\eta$  &  $\delta$ \\
  \hline
  SFCRD   & $1.41(1)$ & $3.32(1)$ & $1.74(2)$ & $0.16(1)$ &
  $0.76(2)$ \\
    Manna & $1.43(1)$ & $3.31(1)$ & $1.75(2)$ & $0.16(2)$
  & $0.75(2)$ 
 \end{tabular}
\caption{Critical exponents for spre\-ading and ste\-ady state
  experiments in $d=3$. Figures in parenthesis indicate the statistical 
  uncertainty in the last digit. Manna exponents from
  Refs.~\protect\cite{rpv00,bigfes,mannasim,mfdp1,mfdp2}. }
\label{tableII}
\end{table}
\newpage

\begin{thebibliography}{10}
\bibitem{notev}
        In spite of the naive power counting analysis, the irrelevance 
        of all terms must be checked on the grounds of a full
        RG analysis.
\bibitem{notecon}
        It is possible to show that the present noise term $\eta_\f$
        is equivalent to a conserved noise; i.e., it generates the same 
        diagrams in a perturbative expansion. After M.A. Mu{\~n}oz
        and F.van Wijland, private communications.
\end{thebibliography}
\end{document}